\documentclass[a4paper]{jps-cp-mod}
\newcommand{\Imm}{\mathop{\rm Im}\nolimits}

\usepackage{txfonts} 

\title{Second Order Perturbation Theory for a Superconducting Double Quantum Dot}

\author{Vladislav \textsc{Pokorn\'{y}}$^{1}$, Martin \textsc{\v{Z}onda}$^{2}$, 
Georgios \textsc{Loukeris}$^{2}$, and Tom\'{a}\v{s} \textsc{Novotn\'{y}}$^{3}$}

\inst{
$^{1}$ Institute of Physics, Czech Academy of Sciences, 
Na Slovance 2, Praha 8 CZ-182~21, Czech Republic\\
$^{2}$Institute of Physics, Albert Ludwig University of Freiburg, 
Hermann-Herder-Strasse 3, Freiburg DE-791~04, Germany\\
$^{3}$ Department of Condensed Matter Physics, 
Faculty of Mathematics and Physics, Charles University,
Ke Karlovu 5, Praha 2 CZ-121~16, Czech Republic}

\email{tno@karlov.mff.cuni.cz}

\recdate{\today}

\abst{
We extend our approach based on the second order perturbation theory in the Coulomb 
interaction recently developed for quantum dots coupled to superconducting leads 
to the superconducting double quantum dot setups. Using our perturbative method 
we evaluate several single-particle quantities such as on-dot induced gap and 
generalized occupations together with the Andreev in-gap spectra and compare 
them with numerically exact results from the Numerical Renormalization Group and 
Quantum Monte Carlo finding a very good correspondence for not too strongly 
correlated regimes. Thus we can offer in a wide parameter range this method as 
an efficient and reliable alternative to the heavy numerical tools exclusively 
used so far for the description of such experimentally relevant systems.   
}

\kword{superconducting quantum dots, diagrammatic perturbation theory, Andreev bound states}

\begin{document}
\maketitle

\section{Introduction}

Large number of experiments involving quantum dots attached to superconductors have been conducted 
in the past two decades~\cite{franceschi2010}. Functional quantum dots in such setups are realized 
using a variety of systems (single molecules, carbon nanotubes, semiconducting nanowires etc.) 
and also various arrangements of several superconducting and/or normal leads exist. Parameters of 
such systems are often tunable, e.g., single-particle energies by the gate voltage and/or 
phase-difference in generalized Josephson junctions by the magnetic flux piercing the SQUID loops, 
which contributes to their versatility. The envisioned applications of such systems range from 
various sensors and detectors (e.g., single-molecule SQUIDs~\cite{cleuziou2006, bouchiat2009}) 
to building blocks of quantum information technologies~\cite{franceschi2010}.

While most of the experimental and theoretical studies have been thus far performed 
for single quantum dots attached to superconducting leads, in past several years the focus 
has partly shifted to double quantum dots (DQDs). Analogously to the single quantum dot setups 
there are many physical realizations of the DQD version using carbon 
nanotubes~\cite{cleuziou2006,pillet2013}, semiconducting 
InAs~\cite{sherman2017,saldana2018,rasmussen2018,saldana2018_2} 
or InSb~\cite{su2017} nanowires, or molecular dimers on 
surfaces~\cite{ruby2018,choi2018,kezilebieke2019}. Apart from their relevance 
for fundamental physics, studying DQD systems is also a first step in understanding 
the behavior of longer nanowires which can host topologically non-trivial 
(Majorana) end states~\cite{su2017} and help to understand the behavior of the 
emerging Andreev bands.      

Theoretical description of DQD setups uses either idealized limiting cases such as the 
superconducting atomic limit \cite{droste2012} assuming very large superconducting gap or 
master equation approach \cite{pfaller2013} with infinitesimally weak coupling to the leads 
(both of these assumptions are not realistic as in experiments the smallest energy scale 
involved is typically the superconducting gap) or heavy numerics such as 
Numerical Renormalization Group (NRG) for realistic experimental parameters~\cite{zitko2015}. 
This is the same situation as in the single dot setups before our discovery that a 
simple second-order perturbation theory in Coulomb interaction gives excellent results 
in a vast (and experimentally relevant) part of the parameter space~\cite{zonda2015,zonda2016}.        

In this work, we extend our second order perturbation theory (2PT) to the DQD setup and 
critically examine its performance by detailed comparison with two numerically exact techniques, 
NRG and the hybridization-expansion quantum Monte Carlo (CT-HYB). We find that, 
analogously to the single dot setups, the 2PT performs very well up to moderately strong 
interaction and can thus be used as a reliable and highly efficient tool for 
semi-quantitative description of DQD systems in a large portion of the parameter space.     

\section{The model Hamiltonian}
\begin{figure}[tbh]
\begin{center}
\includegraphics[width=12cm]{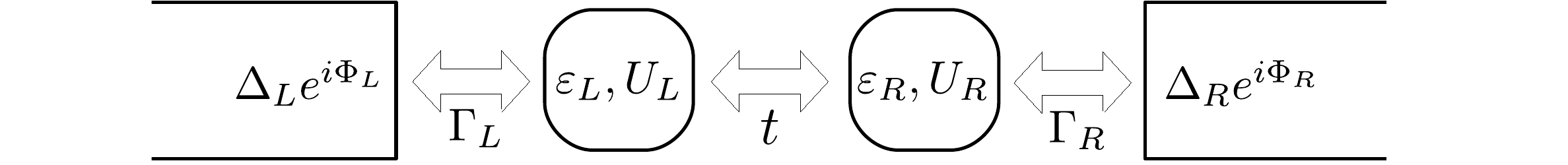}
\caption{Sketch of the serial configuration of DQD connected to superconducting leads.}
\label{Fig:setup}
\end{center}
\end{figure}

The Hamiltonian for the two-impurity Anderson model with two superconducting leads 
in the serial configuration sketched in Fig.~\ref{Fig:setup} where the left lead is 
connected to the left dot and the right lead to the right one reads
\begin{equation}
\mathcal{H}=\mathcal{H}_\mathrm{dots}+\mathcal{H}^{L}_\mathrm{lead}+\mathcal{H}^{R}_\mathrm{lead}
\label{eq:Ham}
\end{equation}
where
\begin{equation}
\mathcal{H}_\mathrm{dots}=\sum_{i\sigma}\varepsilon_{i\sigma} 
d_{i\sigma}^\dag d_{i\sigma}^{\phantom{\dag}}
-\sum_\sigma t_{\sigma}\left(d_{L\sigma}^\dag d_{R\sigma}^{\phantom{\dag}}
+\textrm{H.c.}\right)
+\sum_i U_id_{i\uparrow}^\dag d_{i\uparrow}^{\phantom{\dag}} 
d_{i\downarrow}^\dag d_{i\downarrow}^{\phantom{\dag}}
\label{eq:Dots}
\end{equation}
describes the two QDs and
\begin{equation}
\mathcal{H}^i_\mathrm{lead}=
\sum_{\mathbf{k}\sigma}\varepsilon_{i\mathbf{k}\sigma}
c_{i\mathbf{k}\sigma}^\dag c_{i\mathbf{k}\sigma}^{\phantom{\dag}}
-\Delta_i\sum_\mathbf{k}\left(e^{i\Phi_i}
c_{i\mathbf{k}\uparrow}^\dag c_{i\mathbf{-k}\downarrow}^\dag+\textrm{H.c.}\right)
-\sum_{\mathbf{k}\sigma}V_{i\mathbf{k}\sigma}\left(c_{i\mathbf{k}\sigma}^\dag 
d_{i\sigma}^{\phantom{\dag}}+\textrm{H.c.}\right),\quad i=L,R
\label{eq:Leads}
\end{equation}
is the Hamiltonian describing a BCS superconducting lead and its 
coupling to the quantum dot.
Here $d_{i\sigma}^\dag$ creates an electron on site $i=L,R$ with spin $\sigma$ 
and energy $\varepsilon_{i\sigma}$, $t_{\sigma}$ is the inter-dot hopping amplitude, 
$U_i$ is the local Coulomb interaction (charging energy) on site $i$, 
$c_{i\mathbf{k}\sigma}^\dag$ creates an electron with spin $\sigma$ 
and energy $\varepsilon_{i\mathbf{k}\sigma}$ in lead $i$, 
$\Delta_ie^{i\Phi_i}$ is the complex superconducting order parameter in 
lead $i$ and $V_{i\mathbf{k}\sigma}$ is the hopping between the lead $i$ and the 
corresponding quantum dot. 

For the sake of simplicity we neglect possible capacitive 
coupling $U'n_Ln_R$ between the dots as well as the Rashba coupling 
that is present in semiconductor nanowires~\cite{droste2012}.
From now on we also consider the gap of the same size in both leads, 
$\Delta_L=\Delta_R\equiv\Delta$, as the leads in an experiment are usually 
made from the same material and use $\Delta$ as the energy unit in all results. We also assume 
spin-independent system with  
$t_\uparrow=t_\downarrow\equiv t$,  
$V_{i\mathbf{k}\uparrow}=V_{i\mathbf{k}\downarrow}\equiv V_{i\mathbf{k}}$, 
and $\varepsilon_{i\mathbf{k}\uparrow}=\varepsilon_{i\mathbf{k}\downarrow}\equiv\varepsilon_{i\mathbf{k}}$. 
In all our calculations we use constant tunneling densities of states  
$\Gamma_i(\omega)\equiv\pi\sum_{\mathbf{k}}|V_{i\mathbf{k}}|^2\delta(\omega-\varepsilon_{i\mathbf{k}})=\Gamma_{i}\Theta(D^2-\omega^2)$ 
within the band of finite half-width $D$. 
Unfortunately, we cannot use the symmetry-asymmetry relation from the single quantum dot setup, 
where both leads are connected to the same dot, to map asymmetric coupling 
situations $\Gamma_L\neq\Gamma_R$ onto the symmetric one by a suitable change of the 
phase difference~\cite{kadlecova2017}. Yet, as in any Josephson junction also observables 
in this setup can only depend on the superconducting phase difference 
$\Phi=\Phi_L-\Phi_R$, not on the absolute values of the two phases.

It is often useful to compare the results to the solution in the $\Delta\rightarrow\infty$
superconducting atomic limit~\cite{bauer2007}. 
The Hamiltonian in this limit reads
\begin{equation}
\mathcal{H}_\infty=\mathcal{H}_\mathrm{dots}
+\sum_i\Gamma_i\left(e^{i\Phi_i} 
d_{i\uparrow}^\dag d_{i\downarrow}^\dag+\textrm{H.c.}\right),\quad i=L,R.
\label{eq:HamDinf}
\end{equation}
Its descrete spectrum gives a picture about the structure of the spin multiplets
and helps to better understand the results from numerical methods.

\section{Methods}
\subsection{Second order perturbation theory}
The presented second-order perturbation theory
is a direct generalization of the method introduced 
in Ref.~\cite{zonda2015,zonda2016} for a single superconducting quantum dot.
This method can be straightforwardly applied to 
a non-interacting Green function describing a DQD that is
introduced in the Appendix.
It is based on a diagrammatic expansion in the powers of the Coulomb 
interaction $U$ up to the second order. This simple and fast method provides
a reliable description of the system in the weak and intermediate 
interaction regimes provided the ground state is a singlet, as this method 
fails for degenerate ground states due to the violation of the 
Gell-Mann-Low theorem. However, the calculations performed in 
the $\Delta\rightarrow\infty$ superconducting atomic limit
as well as NRG calculations~\cite{zitko2015} suggest that the ground 
state of a double quantum dot system is a singlet in the vast part of 
the parameter space, making the 2PT method usable in most situations.

The method was implemented in Matsubara (imaginary frequency) formalism using 
the TRIQS libraries~\cite{parcollet2015}. All calculations were 
performed at a small finite temperature $k_BT=10^{-2}\Delta$ with 
a cutoff in imaginary frequencies $\omega_{n}^{max}\geq 500\Delta$. 
The spectral functions 
were obtained by analytic continuation to the real frequency domain,
$G(i\omega_n)\rightarrow G(\omega+i\eta)$ using the 
Pad\'{e} approximation~\cite{vidberg1977}. The continuation was 
performed from the first 50 Matsubara frequencies as adding more 
frequencies did not change the result in any way. A small imaginary 
part $\eta=10^{-3}\Delta$ was added to the real frequency to guarantee 
the correct analytic properties of the continued function.

\subsection{Numerical renormalization group}
We compared our results obtained with 2PT against the zero
temperature data calculated by the 
numerical renormalization group~\cite{bulla2008,hecht2008}, 
using the open source package NRG Ljubljana~\cite{zitko2009,NRG}. 
The logarithmic discretization parameter was set to
$4$, the maximum number of states kept in the truncation was $4000$
(times the multiplicity), the cut-off energy was set to $10$ in the
units of characteristic NRG energy scale and the minimal number of kept
states was set to $1000$.

\subsection{Quantum Monte Carlo}
We also used the TRIQS/CTHYB continuous-time, 
hybridization-expansion quantum Monte Carlo solver \cite{seth2016}
to cross-check the NRG results and assess the effects of 
the finite temperature on the system. Hamiltonian~\eqref{eq:Ham} 
does not conserve the electron number and therefore cannot be solved 
directly using the standard CT-HYB technique. To circumvent this problem 
we utilized a canonical particle-hole transformation in the spin-down 
sector as described in Ref.~\cite{pokorny2018}. The 
continuous-time quantum Monte Carlo is an inherently finite-temperature 
method and all calculations were performed at $k_BT=0.05\Delta$. 
If we consider a typical gap of an InAs nanowire proximitized to Al, 
$\Delta\approx 200\mu$eV~\cite{sherman2017}, this temperature corresponds to 
$T\approx 120$mK. Some of the results were recalculated at 
$k_BT=0.025\Delta$, showing very little temperature dependence of 
the measured quantities.

\section{Results}
All presented results were calculated for a symmetric setup 
$U_L=U_R\equiv U$, $\varepsilon_L=\varepsilon_R\equiv \varepsilon$ 
and $\Gamma_L=\Gamma_R\equiv\Gamma$, for fixed coupling strengths 
$\Gamma=2\Delta$, $t=\Delta$ and zero phase difference $\Phi=0$.
The half-bandwidth of the flat tunneling density of states in the leads is 
set to $D=100\Delta$ in all numerical calculations.

\begin{figure}[tbh]
\begin{center}
\includegraphics[width=7.5cm]{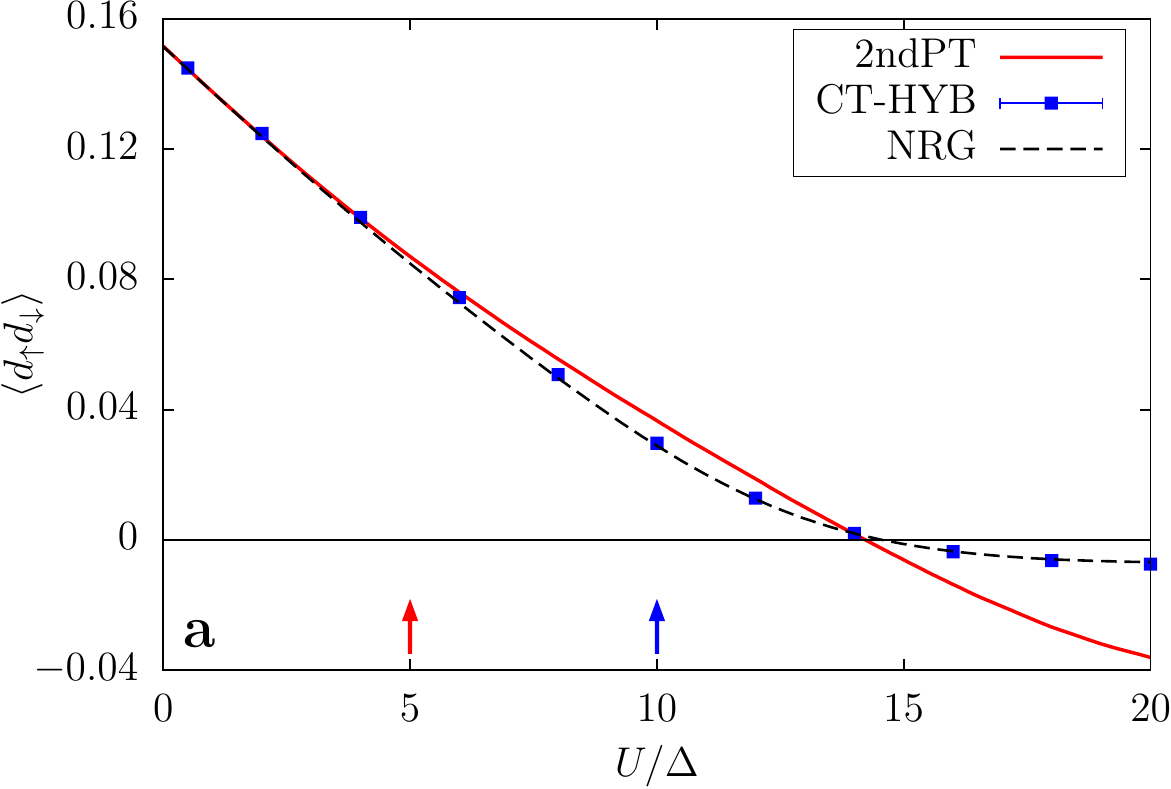}
\includegraphics[width=7.5cm]{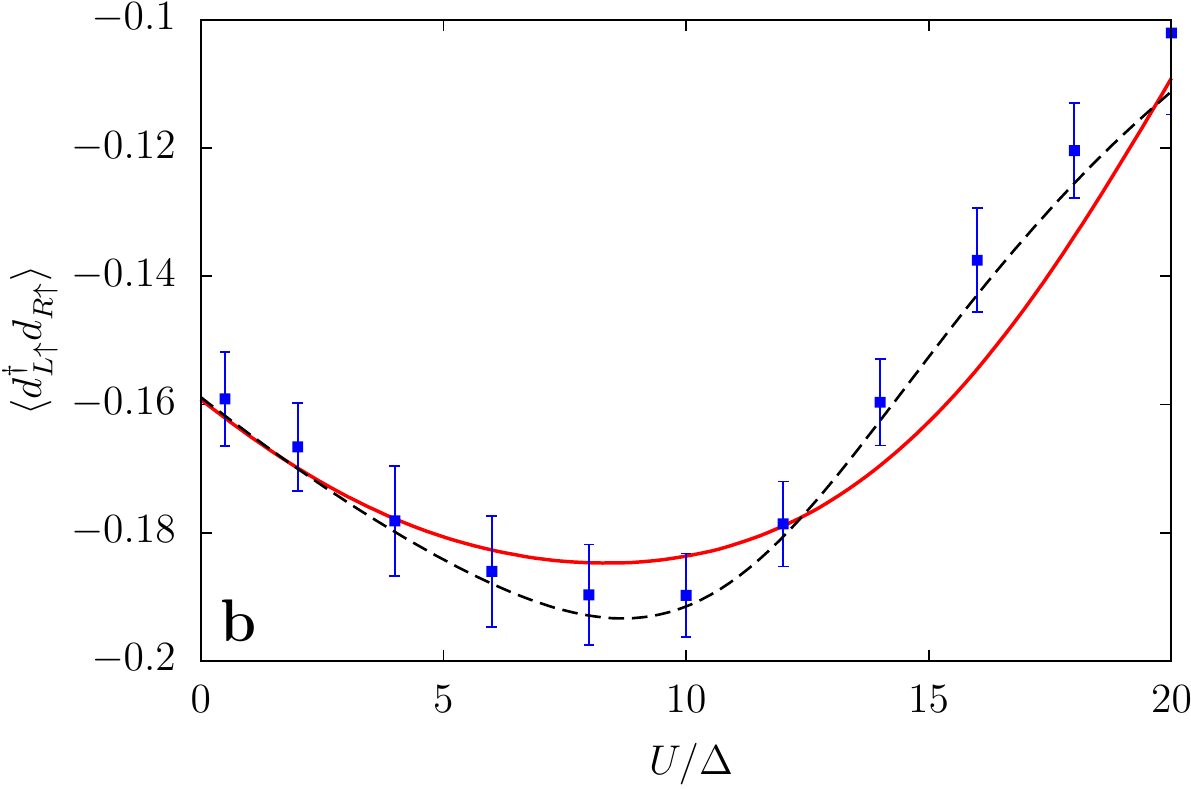}
\caption{Pairing correlation function 
$\nu=\langle d_{L\uparrow} d_{L\downarrow}\rangle$ (panel a)
and the interdot correlation function 
$\lambda=\langle d^\dag_{L\uparrow}d^{\phantom{\dag}}_{R\uparrow}\rangle$ (panel b)
as functions of the interaction strength $U$ at half-filling ($\varepsilon=-U/2$). 
We compared the results of 2PT (red solid line), NRG 
(black dashed line) and CT-HYB (blue squares with error bars).
Parameters are described in the main text. Arrows mark the $U$ values
for which the data in Fig.~\ref{Fig:epsdep} are plotted.}
\label{Fig:Udep1}
\end{center}
\end{figure}

In Fig~\ref{Fig:Udep1} we plotted the dependence of the 
pairing correlation function 
$\nu\equiv\langle d_{L\uparrow} d_{L\downarrow}\rangle
=\langle d_{R\uparrow} d_{R\downarrow}\rangle$ 
(panel a) and the interdot correlation function 
$\lambda\equiv\langle d^\dag_{L\uparrow}d^{\phantom{\dag}}_{R\uparrow}\rangle$ 
(panel b) as functions of the interaction strength $U$ at 
half-filling ($\varepsilon=-U/2$). 
The NRG and CT-HYB results agree within the QMC error bars in the whole
interval of interaction strengths. The 2PT result for $\nu$ agrees with the numerically 
exact solutions up to $U/\Delta\approx 15$ ($U/\Gamma\approx 7.5$). For higher 
values it still correctly describes the qualitative behavior (sign change) of $\nu$. 
The numerical values for $\lambda$ in almost the whole range are less precise, 
yet the overall shape of the curve follows closely the exact numerics including 
the existence and position of its minimum.

\begin{figure}[tbh]
\begin{center}
\includegraphics[width=7.5cm]{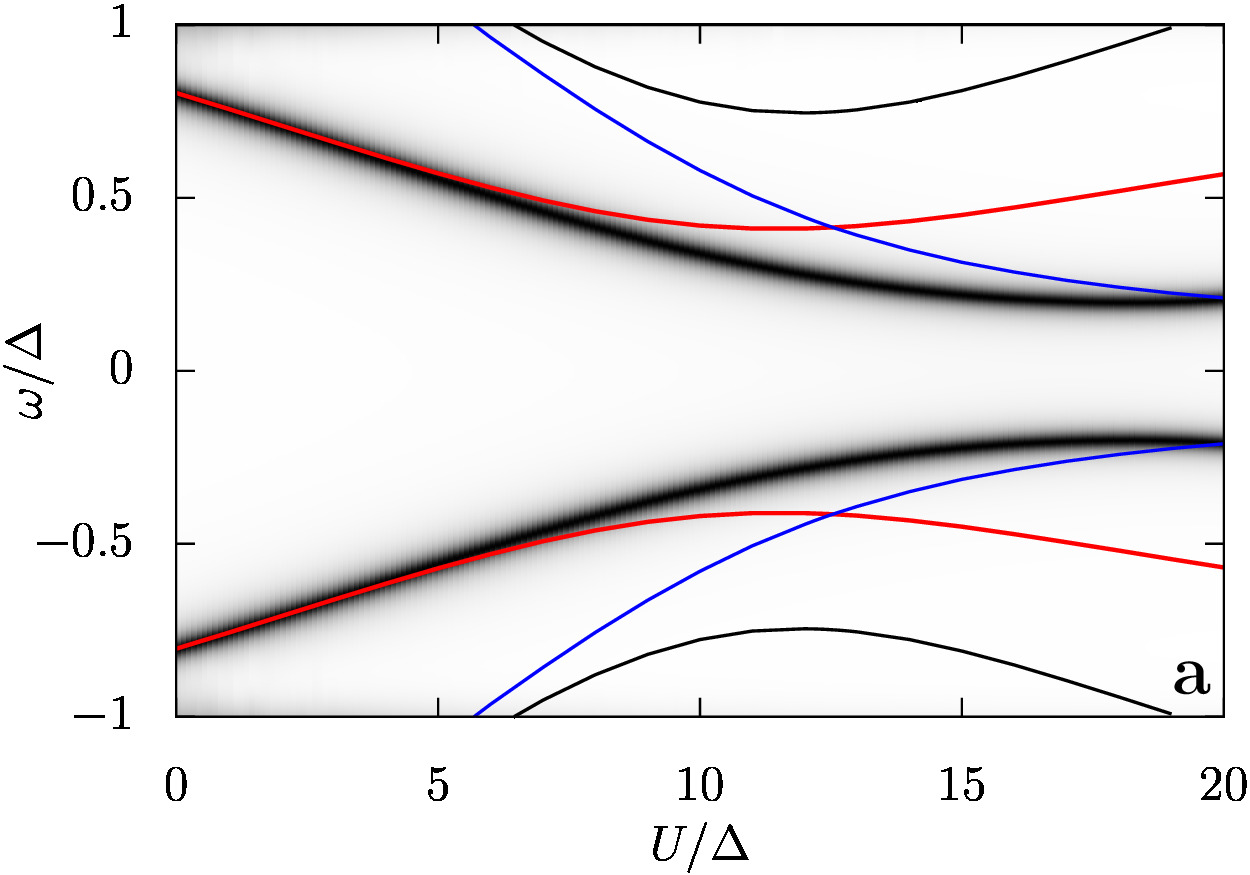}
\includegraphics[width=7.5cm]{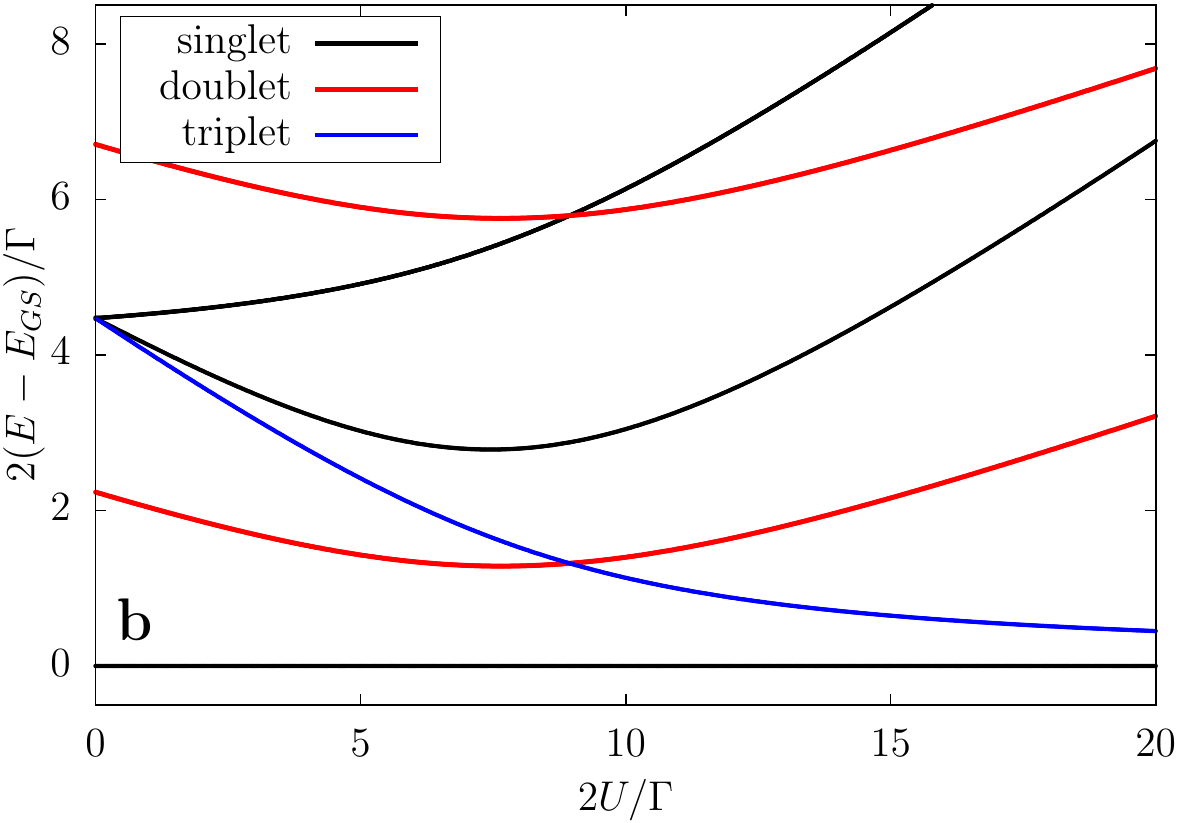}
\caption{Panel a: In-gap excitation spectrum 
for the same parameters as in Fig.~\ref{Fig:Udep1}. 
Color map is the spectral function $\rho(\omega)$ obtained from 2PT result 
using the Pad\'{e} approximation. Lines are NRG results:
singlet-doublet transition (red), singlet-triplet transition (blue), and
singlet-singlet transition (black). 
The latter two are not present in the one-particle spectral function
as they violate the $\Delta S_z=\pm 1/2$ selection rule. 
Panel b: Eigenvalues of the Hamiltonian in the 
$\Delta\rightarrow\infty$ limit w.r.t. the ground state energy $E_{GS}$. 
The axes are scalled by factor of two to be consistent with the left plot as 
$\Gamma=2\Delta$.}
\label{Fig:Udep2}
\end{center}
\end{figure}

In Fig~\ref{Fig:Udep2}a we plotted the in-gap normal spectral 
function $\rho(\omega)\equiv-\Imm G_{11}(\omega+i0)/\pi$ (see Appendix) obtained from the 
imaginary-frequency 2PT solution by analytic continuation to the real axis using
the Pad\'{e} approximation and compared it to the excitation spectrum calculated by NRG. 
Only the singlet-doublet transition (red line) is visible in the 
one-electron spectral function (color map). The singlet-triplet transition 
(blue dashed) and singlet-other singlet transition (black dashed) violate
the $\Delta S_z=1/2$ selection rule and therefore are invisible in the one-electron
spectral function. The 2PT correctly predicts the general behavior of the Andreev bound
states (e.g. the minimum in their energy as function of $U$) and their position for small 
values of $U$, but naturally fails to describe their exact position in the 
strong interaction regime.

In order to better understand the behavior of the Andreev bound states, we also 
calculated the spectrum of the atomic Hamiltonian 
$\mathcal{H}_\infty$~\eqref{eq:HamDinf}
using the atomic solver implemented in
TRIQS and plotted it in Fig~\ref{Fig:Udep2}b. 
We see that the ground state in the wide-band limit is always a singlet 
and the first excited state for small $U$ is a degenerate pair of doublets. 
This degeneracy can be lifted by applying gate voltage $\varepsilon$, 
leading to a splitting of the Andreev bound states (see below).
The second pair of doublets lies high in energies and the transition to them
falls within the continuous band of the finite-$\Delta$ model.

\begin{figure}[tbh]
\begin{center}
\includegraphics[width=7.5cm]{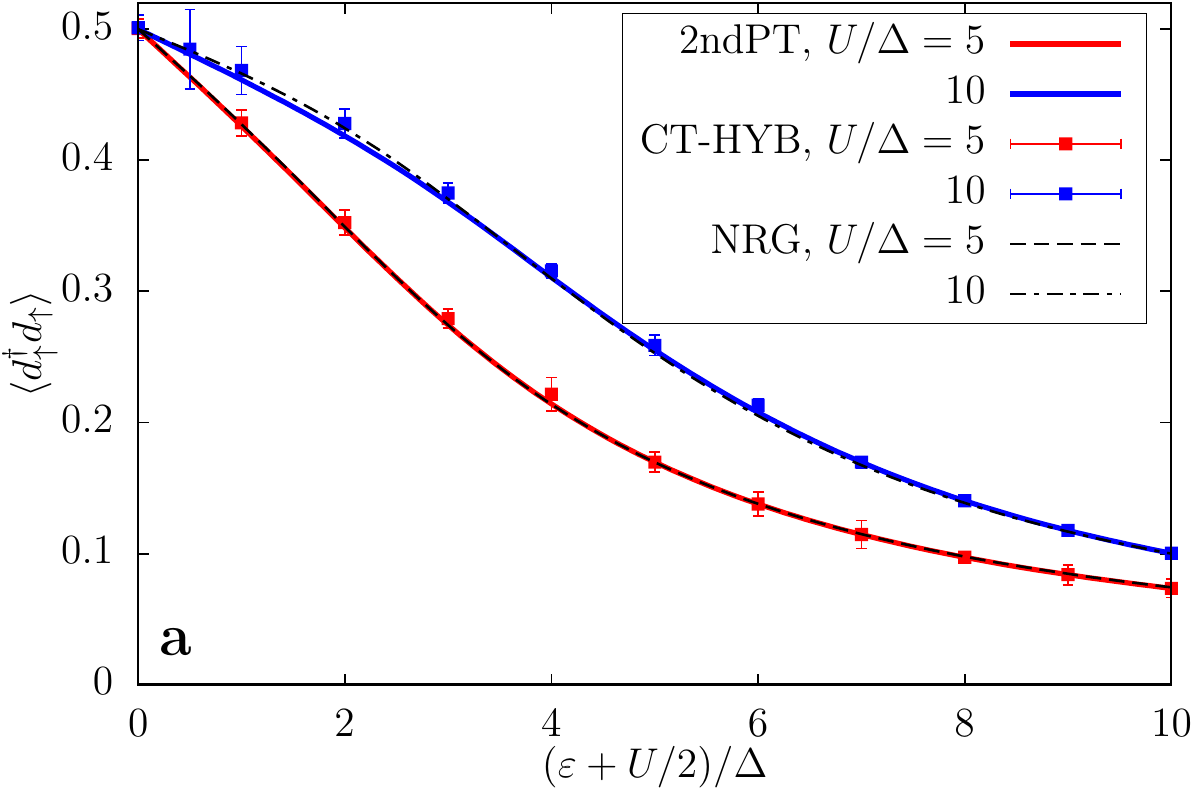}
\includegraphics[width=7.5cm]{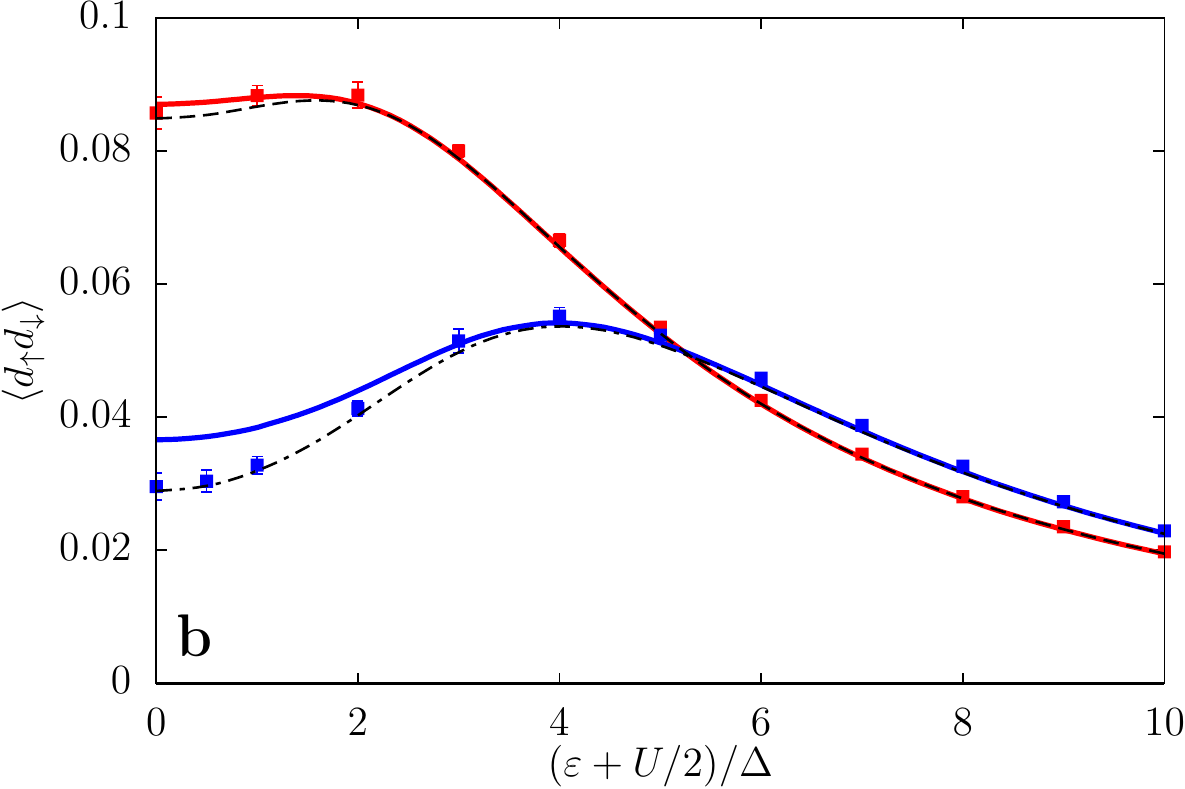} \\
\includegraphics[width=7.5cm]{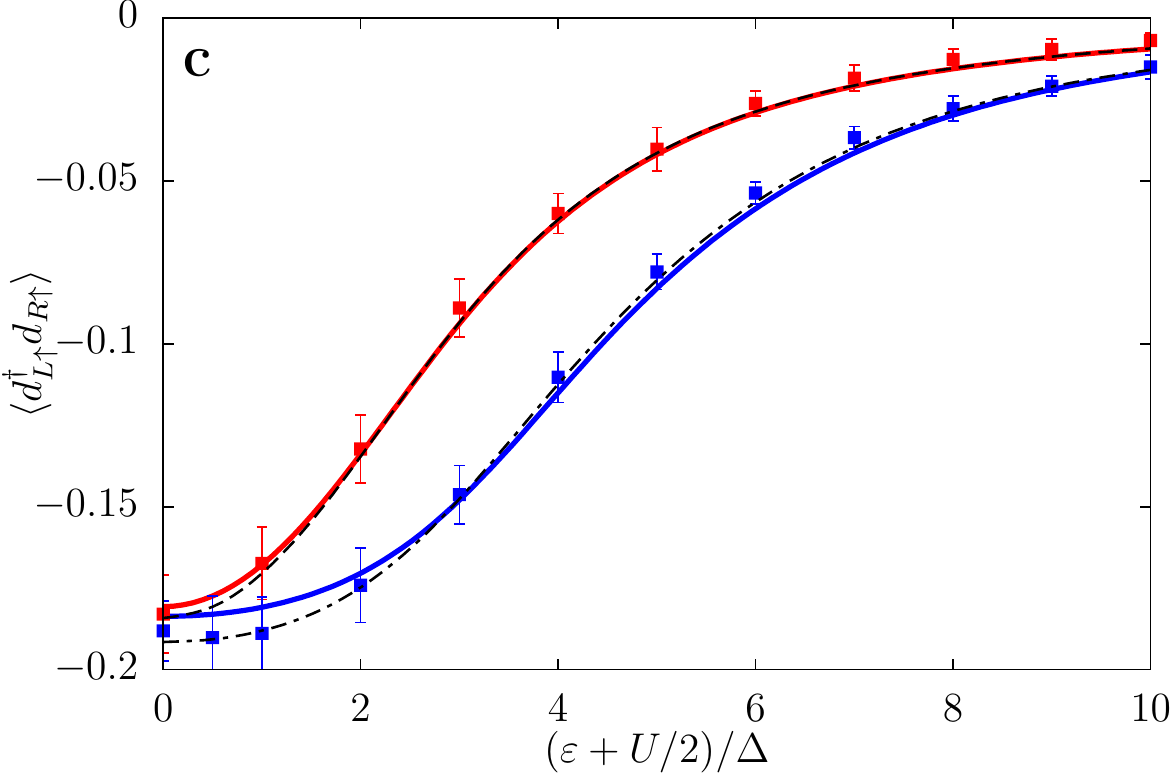}
\includegraphics[width=7.5cm]{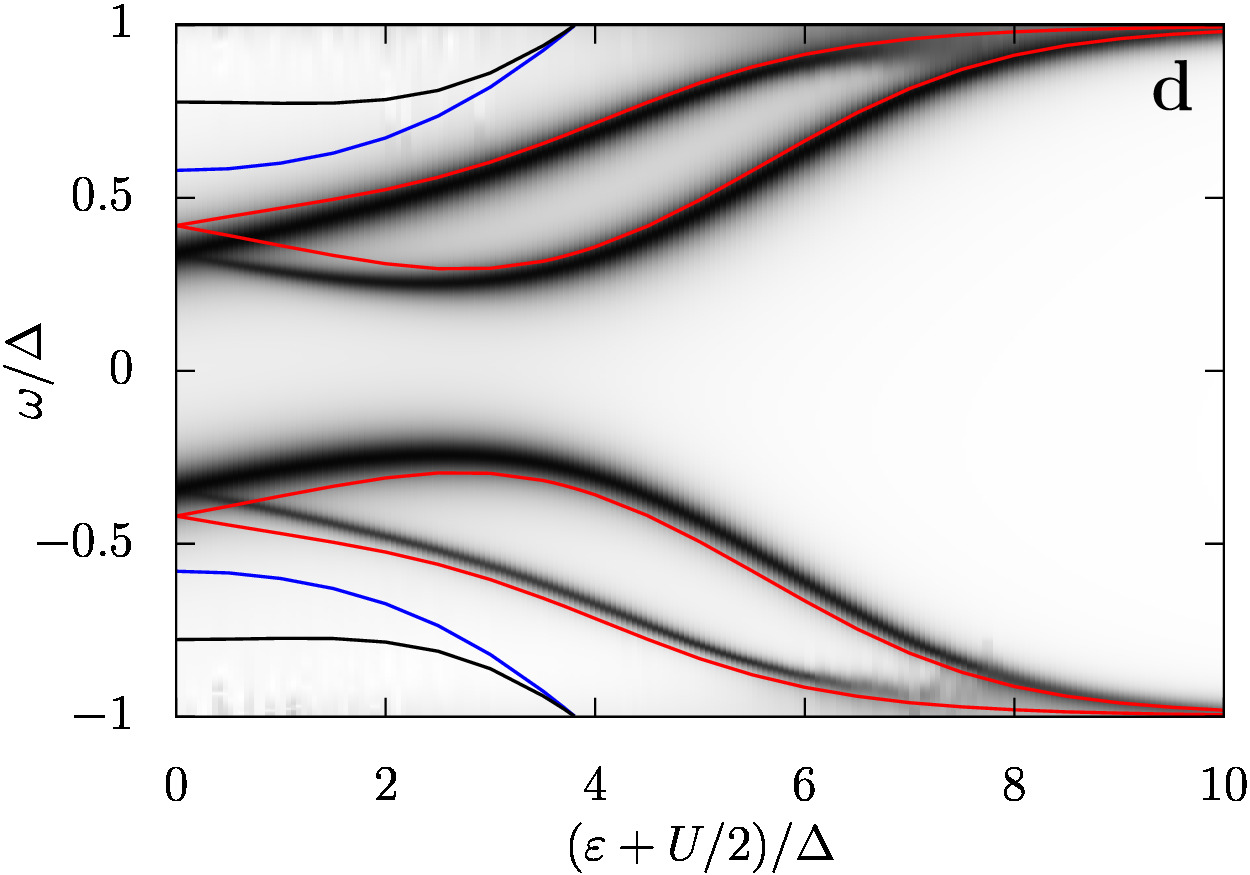}
\caption{Electron density 
$n\equiv\langle d^\dag_{L\uparrow} d^{\phantom{\dag}}_{L\uparrow}\rangle$ (panel a), 
the pairing correlation function 
$\nu\equiv\langle d_{L\uparrow} d_{L\downarrow}\rangle$ (panel b), and
the interdot correlation function 
$\lambda\equiv\langle d^\dag_{L\uparrow}d^{\phantom{\dag}}_{R\uparrow}\rangle$ (panel c)
for two values of the interaction strength as
functions of the local energy 
level $\varepsilon$ w.r.t. half-filling ($\varepsilon=-U/2$).
Panel d: The in-gap excitation spectrum for 
$U=10\Delta$. Color code follows Fig.~\ref{Fig:Udep2}a. 
Noise in the 2PT spectral function $\rho(\omega)$ (color map) 
in the vicinity of the gap edge is an artifact of the Pad\'{e} fitting procedure.}
\label{Fig:epsdep}
\end{center}
\end{figure}

Interaction strength $U$ is a material-dependent quantity and cannot 
be changed easily in the experiment. On the other hand,
local energy level $\varepsilon$ can be tuned by changing the gate 
voltage. In Fig.~\ref{Fig:epsdep} we plotted 
the behavior of a DQD as a function of the energy level for 
two values the interaction strength $U=5\Delta$ and 
$U=10\Delta$ (marked by arrows in Fig~\ref{Fig:Udep1}a). 
The correlation effects get weaker as we move away from 
half-filling ($\varepsilon=-U/2$) because the average dot occupation 
$n\equiv\langle d^\dag_{L\uparrow} d^{\phantom{\dag}}_{L\uparrow}\rangle
=\langle d^\dag_{R\uparrow} d^{\phantom{\dag}}_{R\uparrow}\rangle$ 
(panel a) is decreasing. As a result, 2PT data agree with the 
NRG better as we increase $\varepsilon$.

Panel d shows the in-gap spectral function $\rho(\omega)$.
The ground state is again a singlet and we see two pairs of Andreev bound states
corresponding to the singlet-doublet transitions. 
This is a result of the splitting of the two degenerate doublets by the 
applied gate voltage mentioned earlier in the text. As the average dot
occupation is decreasing, the Andreev bound states move closer to the gap edges, 
eventually merging into the continuum, leaving the spectral function 
free of any in-gap states. The singlet-triplet (blue) and singlet-other singlet (black)
transitions are again invisible to the one-particle spectral function
as mentioned in the previous text and Fig.~\ref{Fig:Udep2}a.

\section{Conclusions}
We presented a fast and simple perturbation method to calculate 
properties of a DQD system with superconducting leads and benchmarked 
it against exact numerical techniques as the numerical renormalization group
and the quantum Monte Carlo. It is usable in the weak and intermediate 
correlation regimes, provided that the ground state is a singlet, which 
is true in most situations. In the strongly correlated regime
it can still provide qualitatively correct description of the system. 
The main limitation of this approach lies in its inability (so far) to describe phases 
with degenerate ground states (doublet or triplet in DQD) and, consequently, also to 
provide finite temperature results close to phase transitions where the adjacent phases 
thermally mix~\cite{kadlecova2019}.     
On the other hand, this method can be straightforwardly generalized to more complicated
setups with various geometries, metallic leads and the inter-dot capacitive coupling.

\section*{Acknowledgments}
This work was supported by Grant INTER-COST LTC19045 (V.~P.) and Grant
No. 19-13525S of the Czech Science Foundation (T.~N.). 
Computational resources are provided by The Ministry of Education, 
Youth and Sports from the Large Infrastructures
for Research, Experimental Development and Innovations project 
“IT4Innovations National Supercomputing Center - LM2015070.”

\appendix
\section{The non-interacting Green function}
The input to both the 2PT solver and the CT-HYB solver is the non-interacting 
($U=0$) Green function. We define a Nambu spinor 
$\Psi=\left(d_{L\uparrow}^{\phantom{\dag}},d_{L\downarrow}^{\dag},
d_{R\uparrow}^{\phantom{\dag}},d_{R\downarrow}^{\dag}\right)$
for the DQD. The non-interacting, imaginary-time Nambu-Green function
$\hat{G}_0(\tau)=-\langle \mathcal{T}_\tau[\Psi(\tau)\Psi^\dag(0)]\rangle$ 
is then a $4\times 4$ matrix,
\begin{equation}
\hat{G}_0(\tau)=-\begin{pmatrix}
\langle d_{L\uparrow}^{\phantom{\dag}}d_{L\uparrow}^\dag\rangle_\tau & 
\langle d_{L\uparrow}^{\phantom{\dag}}d_{L\downarrow}^{\phantom{\dag}}\rangle_\tau &
\langle d_{L\uparrow}^{\phantom{\dag}}d_{R\uparrow}^\dag\rangle_\tau & 
\langle d_{L\uparrow}^{\phantom{\dag}}d_{R\downarrow}^{\phantom{\dag}}\rangle_\tau \\[0.1em]
\langle d_{L\downarrow}^\dag d_{L\uparrow}^\dag \rangle_\tau & 
\langle d_{L\downarrow}^\dag d_{L\downarrow}^{\phantom{\dag}}\rangle_\tau & 
\langle d_{L\downarrow}^\dag d_{R\uparrow}^\dag\rangle_\tau & 
\langle d_{L\downarrow}^\dag d_{R\downarrow}^{\phantom{\dag}}\rangle_\tau \\[0.1em]
\langle d_{R\uparrow}^{\phantom{\dag}}d_{L\uparrow}^\dag\rangle_\tau & 
\langle d_{R\uparrow}^{\phantom{\dag}}d_{L\downarrow}^{\phantom{\dag}}\rangle_\tau & 
\langle d_{R\uparrow}^{\phantom{\dag}}d_{R\uparrow}^\dag\rangle_\tau & 
\langle d_{R\uparrow}^{\phantom{\dag}}d_{R\downarrow}^{\phantom{\dag}}\rangle_\tau \\[0.1em]
\langle d_{R\downarrow}^{\dag}d_{L\uparrow}^\dag\rangle_\tau & 
\langle d_{R\downarrow}^{\dag}d_{L\downarrow}^{\phantom{\dag}}\rangle_\tau & 
\langle d_{R\downarrow}^{\dag}d_{R\uparrow}^\dag\rangle_\tau & 
\langle d_{R\downarrow}^{\dag}d_{R\downarrow}^{\phantom{\dag}}\rangle_\tau \\
\end{pmatrix}
\end{equation}
where we denoted $\langle xy\rangle_\tau=
\langle \mathcal{T}_\tau[x(\tau)y(0)]\rangle$.
The Green function in the Matsubara frequency domain reads
\begin{equation}
\hat{G}_0(i\omega_n)=\int_0^\beta d\tau e^{i\omega_n\tau}\hat{G}_0(\tau)
=\left[i\omega_n\hat{I}-\hat{\varepsilon}-\hat{\Gamma}(i\omega_n)\right]^{-1}
\end{equation}
where $\omega_n=(2n+1)\pi k_BT$, $\hat{I}$ is a $4\times 4$ unit matrix,
\begin{equation}
\hat{\varepsilon}=
\begin{pmatrix}
\varepsilon_{L\uparrow} & 0 & -t_\uparrow & 0 \\[0.1em]
0 & -\varepsilon_{L\downarrow} & 0 & t_\downarrow \\[0.1em]
-t_\uparrow & 0 & \varepsilon_{R\uparrow} & 0 \\[0.1em]
0 & t_\downarrow & 0 & -\varepsilon_{R\downarrow} \\
\end{pmatrix},\qquad
\hat{\Gamma}(i\omega_n)=
\begin{pmatrix}
~\hat{\Gamma}_L(i\omega_n) & \hat{0} \\[0.1em]
\hat{0} & \hat{\Gamma}_R(i\omega_n)~
\end{pmatrix}
\end{equation}
and 
\begin{equation}
\hat{\Gamma}_i(i\omega_n)=
\frac{\Gamma_i}{\sqrt{\Delta^2+\omega_n^2}}
\frac{2}{\pi}\arctan\left(\frac{D}{\sqrt{\Delta^2+\omega_n^2}}\right)
\begin{pmatrix}
i\omega_n & \Delta e^{i\Phi_i}\\[0.1em]
\Delta e^{-i\Phi_i} & i\omega_n
\end{pmatrix},\quad i=L,R.
\end{equation}
Here $\hat{\varepsilon}$ describes the local energy levels and hoppings 
in the isolated DQD and $\hat{\Gamma}_i(i\omega_n)$
is the hybridization function describing the coupling between the 
quantum dot $i=L,R$ and the corresponding superconducting lead
with $2/\pi\arctan\left(D/\sqrt{\Delta^2+\omega_n^2}\right)$ being a correction due to finite bandwidth $D$.
The interacting Green function is a solution of a matrix Dyson equation 
$\hat{G}(i\omega_n)^{-1}=\hat{G}_0(i\omega_n)^{-1}-\hat{\Sigma}(i\omega_n)$, 
with the interaction self-energy matrix $\hat{\Sigma}(i\omega_n)$ calculated by 
the procedure described in Refs.~\cite{zonda2015,zonda2016}.

\end{document}